\documentclass[prl,aps,twocolumn,showpacs,amsmath,amssymb,superscriptaddress]{revtex4}

\usepackage{graphicx}

\usepackage{dcolumn}

\usepackage{bm}

\usepackage[latin1]{inputenc}

\usepackage[T1]{fontenc}

\usepackage{floatflt}

\usepackage{hyperref}

\usepackage{natbib}

\begin{document}



\title{State preparation and dynamics of ultracold atoms in higher lattice orbitals}

\author{Torben Müller}
\affiliation{Institut für Physik, Johannes Gutenberg-Universität, Staudingerweg 7, 55118 Mainz, Germany}
\affiliation{Institute of Quantum Electronics, ETH Zürich, Hönggerberg, CH-8093 Zürich, Switzerland}
\author{Simon Fölling}
\author{Artur Widera}
\affiliation{Institut für Physik, Johannes Gutenberg-Universität, Staudingerweg 7, 55118 Mainz, Germany}
\author{Immanuel Bloch}   
\email{bloch@uni-mainz.de}  
\affiliation{Institut für Physik, Johannes Gutenberg-Universität, Staudingerweg 7, 55118 Mainz, Germany}

\date{\today}


\begin{abstract}
We report on the realization of a multi-orbital system with ultracold atoms in the excited bands of a 3D optical lattice by selectively controlling the band population along a given lattice direction. The lifetime of the atoms in the excited band is found to be considerably longer (10-100 times) than the characteristic time scale for inter-site tunneling, thus opening the path for orbital selective many-body physics with ultracold atoms. Upon exciting the atoms from an initial lowest band Mott insulating state to higher lying bands, we observe the dynamical emergence of coherence in 1D (and 2D), compatible with Bose-Einstein condensation to a non-zero momentum state.
\end{abstract}



\maketitle
Ultracold neutral atoms in optical lattices form a remarkable model system for investigations of strongly correlated quantum phases, highly relevant to modern solid state physics \cite{jaksch05a,bloch05a}.
So far, experiments have mostly concentrated on strongly interacting atoms in the lowest energy band of an optical lattice. In solid state physics, however, orbital selective phenomena are known to play a crucial role for the description of topical materials with strong electronic correlations, such as the transition-metal oxides exhibiting metal-insulator transitions, superconductivity and colossal magnetoresistance.  The extension to excited Bloch bands with ultracold atoms in optical lattices allows to enter a regime beyond $s$-wave isotropy of the on-site atomic wave function on the one hand, and promises a novel route for the realization of long range interactions on the other hand. Recently, these aspects have attracted much attention and have lead to a series of predictions for novel effects and quantum phases that could be realized with ultracold atoms in higher-lying orbitals. These include the generation of novel multi-flavor and multi-orbital systems \cite{Girvin05,Kuklov06,Xu07},  supersolid quantum phases in cubic lattices \cite{Scarola05,Scarola06}, quantum stripe ordering in triangular lattices \cite{Wu06} or Wigner crystallization in honeycomb lattices \cite{Wu07}. 
An essential experimental question is however, how such systems could be prepared and whether they are stable.

In this Letter, we present the experimental realization of a $p$-wave orbital system in the first excited Bloch band of a 3D optical lattice. By driving  stimulated Raman transitions \cite{Wineland03} between the different energy bands, starting from an initially Mott insulating state, we are able to  control the population in different Bloch bands. Previously it had been shown that for an array of quasi 2D degenerate Bose-gases, such excitations tend to decay very rapidly \cite{Spielmann06}. However, in the 3D lattice configuration, we find that such decay processes can be significantly suppressed, due to a reduced final density of states for inelastic scattering events \cite{Girvin05}. Here, the lifetimes of the atoms in $p$-wave orbitals can extend up to hundred tunneling times, thus opening the avenue for orbital selective physics with ultracold atoms. Starting from an initially Mott insulating state in the lowest energy band, we transfer the population in one or two dimensions into excited $p$-wave orbitals.  We find that upon quenching the system into the superfluid regime of the excited bands, it exhibits a dynamical evolution to a state with long range coherence at non-zero quasi-momentum, providing a possible route for Bose-Einstein condensation to non-zero momentum \cite{Liu06}.
 \begin{figure}
	\begin{center}
	\includegraphics[width=0.43\textwidth]{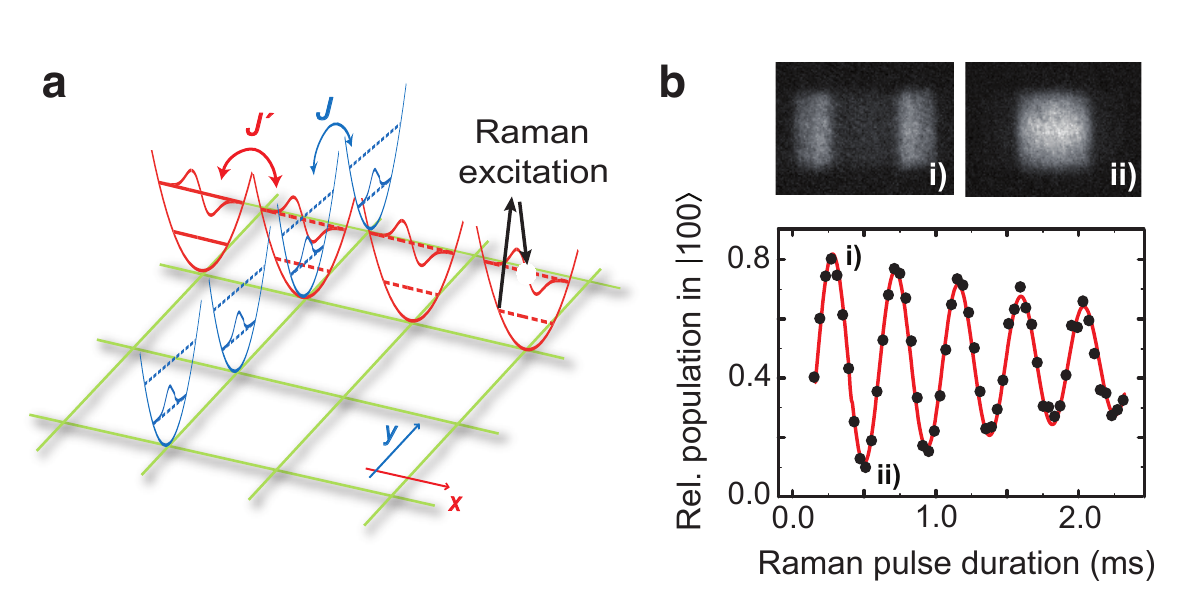}
\end{center}
\caption[]{{\bf (a)} Schematic of anisotropic tunneling following an excitation along the $x-$direction into a $p$-wave orbital using stimulated Raman transitions.  {\bf (b)} Rabi oscillations between the $s-$ and $p-$wave orbital demonstrate the coherent coupling between the levels. }
	\label{fig:RamanProcess}
\end{figure}

We prepare our sample by first loading a quasi-pure BEC of typically $2 \times 10^{5}$ $^{87}$Rb atoms into a deep, simple cubic 3D optical lattice formed by three far detuned standing wave light fields along the $x-,y-,$ and $z-$directions. Here, the lattice potential can be approximated by an array of separated 3D harmonic oscillators with discrete vibrational levels. We denote the transition frequencies along one axis, e.g.~along the $x$-axis, between the vibrational levels $|n_{x}n_{y}n_{z}\rangle$ and $|n'_{x}n_{y}n_{z}\rangle$ as $\omega^{(x)}_{n_x^{\phantom{\prime}} n_x^\prime}$, where $n_j$ is the vibrational quantum number along the $j$-axis. Population can be coherently transferred between these levels using a stimulated two-photon Raman process between the initial and final external states with an effective Rabi frequency $\Omega_{\mathrm{eff}} =  \frac{\Omega_{1}\Omega_{2}^{*}}{2\Delta}\cdot \left|\left\langle n'_{x}n_{y}n_{z} \left| \ e^{i(\mathbf{k}_{L_{1}}-\mathbf{k}_{L_{2}})\cdot\mathbf{x}} \ \right|n_{x}n_{y}n_{z}\right\rangle\right|^{2}$. Here, $\Omega_{1}$  ($\Omega_{2}$) denotes the resonant Rabi frequency for the single-photon transition on the $^{87}$Rb D$_{2}$-line to which the two counter-propagating Raman beams with wave vectors $\mathbf{k}_{L_{1}}$ and $\mathbf{k}_{L_{2}}$ along the $x$-axis are off-resonant by the detuning $\Delta$.

For the experiments presented here, we want to restrict the Raman coupling to the two lowest Bloch bands along the $x$-direction only. Therefore, we initially configure the lattice along the orthogonal $y-$ and $z-$ lattice directions at a large lattice depth of $V_y = V_z = 55\,E_{r}$, where $E_{r}$ is the single-photon recoil energy $E_{r} = h^{2}/2m\lambda_\mathrm{lat}^2$ with $\lambda_\mathrm{lat}$ = 843\,nm the lattice laser wavelength and $m$ the mass of a single atom. The $x-$axis is initially prepared at $V_x= 40\,E_{r}$, corresponding to an excitation frequency of $\omega^{(x)}_{01}\approx 2\pi\times 40$\,kHz. While loading the atoms into this lattice configuration, we avoid any population of higher bands both from heating processes during the ramp and from interaction-induced coupling \cite{Koehl2005a,Chin06} by ensuring that the inverse ramp time and the on-site interaction is much smaller than the band gap. Due to the anharmonicity of the trapping potential at a lattice site, the excited vibrational states are not equally spaced. Therefore, it is possible to avoid additional coupling of the first to the second excited state $\left|200\right\rangle$ if the coupling strength $\Omega_\mathrm{eff}$ fulfills $\Omega_{\mathrm{eff}} \ll \omega^{(x)}_{01} - \omega^{(x)}_{12} \approx 2\pi\times 4.6\,$kHz for our lattice parameters. The coupling strength should however also be large compared to a broadening of the level spacing caused by the tunnel coupling and spatial inhomogeneities due to the Gaussian beam profile of the lattice lasers. Then the stimulated Raman process can drive efficient coherent Rabi-oscillations between the two states $\left|000\right\rangle$ and $\left|100\right\rangle$, without significant population of any other excited vibrational state  (see Fig.~\ref{fig:RamanProcess}b). For a beam waist of approximately $150\,\mu$m of the Raman laser and a detuning of $\Delta = 2 \pi \times 6.8\,$GHz from the $^{87}$Rb D$_{2}$-line, we find typical optimal coupling strengths of $\Omega_\mathrm{eff}\approx 2\pi\times$ 2.2\,kHz. The Rabi oscillations are damped on a time scale of 2.5\,ms which coincides with the calculated inhomogeneity of 400\,Hz due to the harmonic confinement across the atom cloud. From these Rabi oscillations, we deduce the pulse length of a Raman $\pi$-pulse for the $\left|000\right\rangle \leftrightarrow \left|100\right\rangle$ transition for which we achieve a transfer efficiency of nearly 80\%.

The atomic sample is detected using absorption imaging after a ballistic expansion. When the trap is abruptly switched off before this time-of-flight (TOF) period, the momentum distribution after TOF approximatively represents the Fourier transform of the original macroscopic wave function in the lattice. 
 Alternatively, the population of each vibrational (orbital) band can be measured by mapping it to the corresponding Brillouin zone \cite{Greiner01f, Kastberg95, Koehl2005a} when the lattice potential is ramped down adiabatically with respect to the on-site trapping frequency (images in Fig.~\ref{fig:RamanProcess}b).


The lifetime of the atoms in excited bands is a crucial point for the realization of many proposed novel quantum phases.
In general, the population in the first excited band constitutes a meta stable state that can decay by collisional events \cite{Kastberg95}. In a 3D optical lattice, however, energy conservation combined with a small number of accessible final scattering states suggests that decay processes are suppressed. In particular, the anharmonic spacing of the vibrational levels should inhibit the first order decay process, which transfers two colliding $p$-wave bosons into one higher excited second band orbital and one ground state orbital. A more detailed analysis shows that higher order processes, i.e.~involving more particles and higher vibrational states, can lead to a decay of the excited population \cite{Girvin05}.
\begin{figure}
	\begin{center}
	\includegraphics[width=0.36\textwidth]{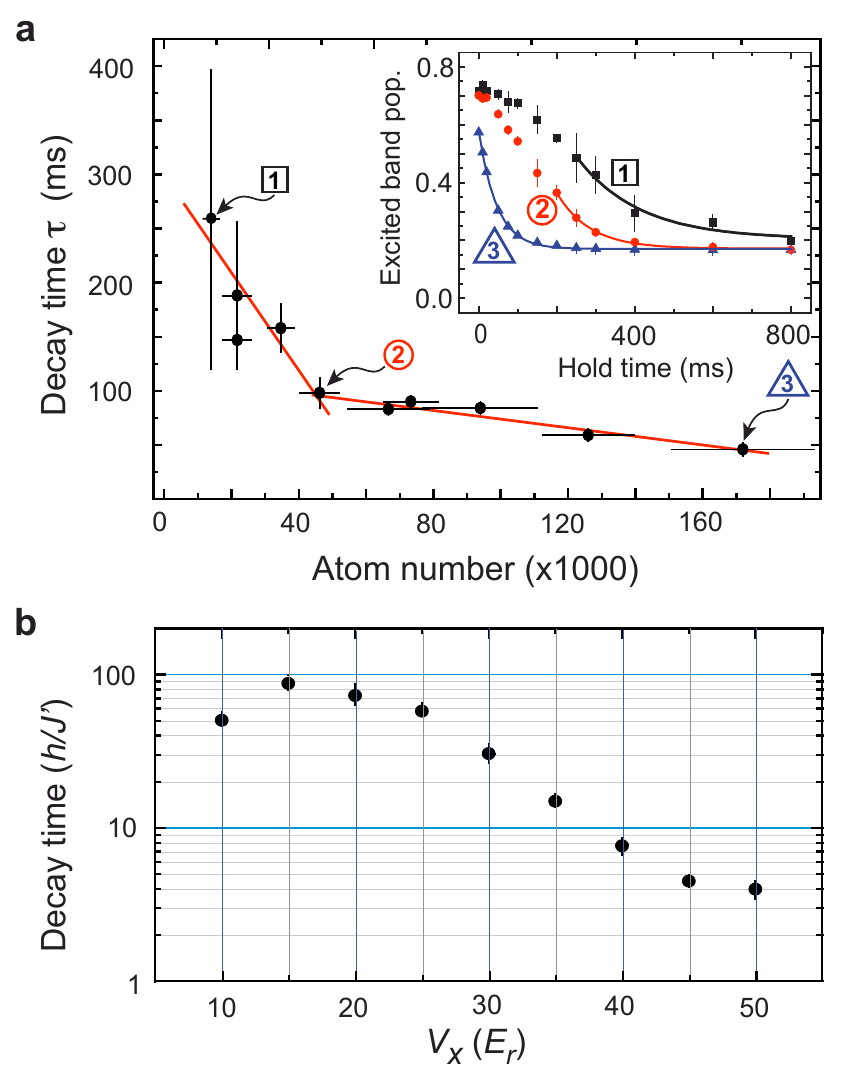}
\end{center}
\caption[]{Measured lifetime of the excited state $\left|100\right\rangle$ with respect to {\bf (a)} the total atom number for $V_x=40\,E_{r}$ and {\bf (b)} the lattice depth $V_x$ along the excitation axis (for 4.5$\times 10^{4}$ atoms). The lifetime $\tau$ was determined from an exponential decay fit to the measured population dynamics $\propto \rm{e}^{-t/\tau}$ (see inset) neglecting short time effects.}
\label{fig:Lifetime}
\end{figure}

In order to experimentally determine the lifetime of the first excited band, we measure the population in the different vibrational states versus the hold time in the lattice for different atom numbers and various lattice depths along the excitation axis. 
For the measurement versus atom number, we first prepare the system in the first excited state $|100\rangle$ by a Raman $\pi$-pulse. We subsequently record the populations of ground and excited vibrational states over time.
For large atom numbers, we observe an immediate start of the decay of population from the first excited band, which is well described by an exponential decay of the form $\exp(-t/\tau)$ (see inset in Fig.~\ref{fig:Lifetime}a).
For decreasing atom number, the decay is delayed for short times. This can be understood through the atom number distribution in the lattice. For large atom numbers, many lattice sites are expected to be occupied by more than one atom. Consequently, the atoms can interact without delay. For decreasing atom number, less and less sites are occupied by more than one atom, until only isolated atoms are trapped in the lattice. In this case, atoms have to tunnel before they collide with another atom and subsequently decay.
Excluding the short time behavior, we also fit the decay for small atom numbers with an exponential decay. In Fig.~\ref{fig:Lifetime}a we plot the decay time $\tau$ versus the initial atom number. The data shows a strong atom number dependence, with decreased lifetimes for larger atom numbers.  Moreover, the data reveals a kink in the decay time for atom numbers of approximately 4.5$\times 10^{4}$, which roughly matches the point where in our system we expect a large fraction of doubly occupied sites to occur \cite{Foelling06}. Below this atom number we observe a strongly rising lifetime of the system up to a factor of five compared to the higher density case, compatible with the interpretation that several tunneling times are required for collisions to occur. 


This interpretation is supported by an additional measurement of the lifetime versus lattice depth.
Lowering the lattice depth implies increased tunneling rates which enhances the rate of atom-atom collisions, thus leading to reduced lifetimes. Furthermore, due to the larger bandwidth, the suppression of inelastic collisions, resulting from the anharmonicity of the potential, is reduced. Fig.~\ref{fig:Lifetime}b shows the measured lifetime of $\left|100\right\rangle$  for different lattice depths along the excitation axis, normalized to the corresponding tunneling time scale $h/J^\prime$. For all these measurements, the sample was prepared in the same way at an $x$-axis lattice depth of 40\,$E_{r}$ before this axis was ramped to the final value in 50\,$\rm{\mu}$s. 
Interestingly, the lifetime of the first excited Bloch band can exceed the time scale associated with inter-site hopping dynamics by a factor of almost 100. The lattice depth-dependence of this scaling shows a similar behavior as recently calculated lifetimes in the first excited band, which accounts for for higher order decay processes \cite{Girvin05}.

\begin{figure}[ht]
	\begin{center}
	\includegraphics[width=\columnwidth]{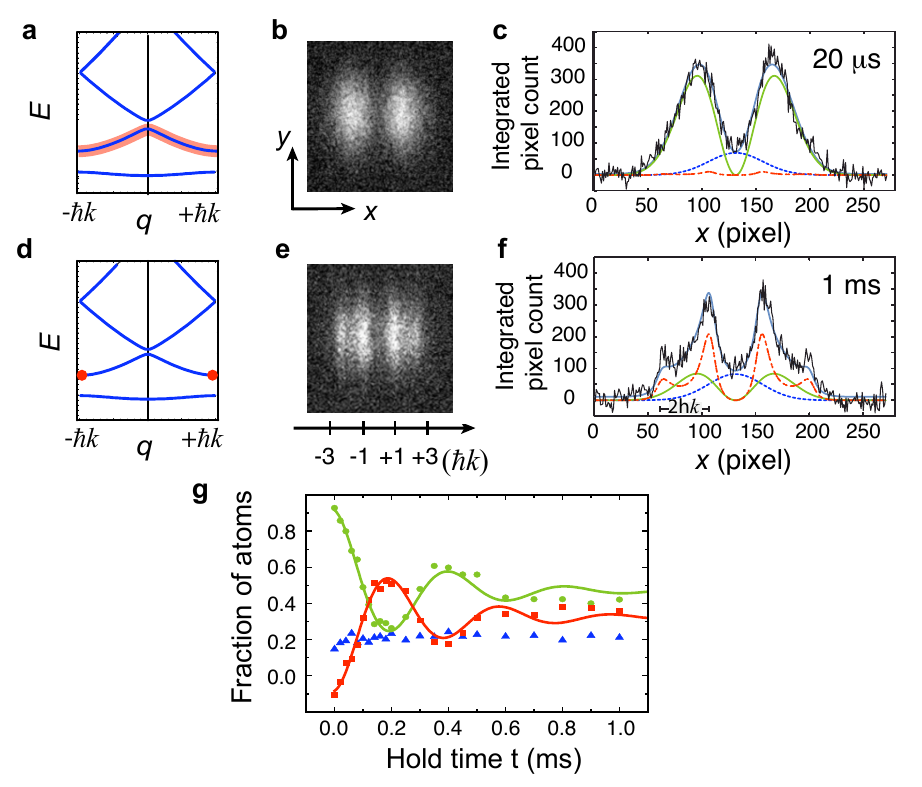}
\end{center}
\caption[]{Emergence of 1D coherence in the first excited Bloch band:
Directly after lowering the latttice depth to $V_{hold}=17\,E_r$, the first excited band is still homogeneously populated {\bf (a)} and the TOF image shows the Wannier function of the first excited state as envelope {\bf (b)}. Within 1\,ms hold time, the system relaxes to a non-zero momentum state {\bf (d)} and shows interference structure {\bf (e)}. {\bf (c,f)} Corresponding horizontal integrated profiles with fits to the coherent contribution of the first excited state (red) and incoherent contributions of the vibrational ground state (blue) and first excited band (green). {\bf (g)} Time evolution of the coherent and incoherent contributions at $V_{hold}=20\,E_r$ (color code as above).}
\label{fig:1D_coherence}
\end{figure}

The long lifetimes of the excited state $\left|100\right\rangle$ compared to the tunneling time scale should allow for the atoms to delocalize and exhibit spatial coherence within the first excited Bloch band when the system is quenched into the superfluid regime for this band. 
In order to observe the emergence of coherence, we excite our initially prepared Mott-insulator by a Raman $\pi$-pulse into the first excited band along the $x$-axis. Subsequently, we reduce the lattice depth in the excitation axis within 50\,$\rm{\mu}$s to a lattice depth $V_{hold}$ less than 25\,$E_{r}$. There we expect the lowest energy state in the first excited band to be superfluid. The corresponding in-trap momentum distribution is then recorded after a TOF period for varying hold times by switching off all trapping potentials. Right after lowering the lattice depth the atoms are still localized at individual lattice sites, corresponding to a homogeneously filled excited band (see  Fig.~\ref{fig:1D_coherence}a). In this situation, the system features no well-defined phase relation between adjacent lattice sites and consequently does not show any interference pattern but the corresponding on-site momentum Wannier function of the state $\left|100\right\rangle$ (see  Fig.~\ref{fig:1D_coherence}b). For hold times longer than the tunneling time at the corresponding lattice depth, tunneling processes and collisions lead to the relaxation of the population in the first Bloch band toward the points of lowest energy, i.e.~the edges of the dispersion relation, see Fig.~\ref{fig:1D_coherence}d. These points correspond to a well-defined phase shift of $\Delta\phi = \pi$ between the atomic wavefunction on neighboring lattice sites leading to the observation of interference fringes in TOF for odd multiples of $\pm\hbar \mathbf{k} $ along the $x$-axis (see  Fig.~\ref{fig:1D_coherence}e), with $\mathbf{k}$ being the wavevector of the lattice laser light. Interestingly, this situation implies the relaxation to a non-zero momentum state of the system \cite{Liu06}.
For hold times longer than 3\,ms, the steady decay of the first excited vibrational level and the increasing population of the vibrational ground state lead to an observed loss of interference contrast on a time scale which corresponds to decay processes as discussed before. In order to quantify our observation, we first integrate the measured absorption images (Fig.~\ref{fig:1D_coherence}b,e) along the $y$-axis. We fit the resulting profiles (Fig.~\ref{fig:1D_coherence}c,f)
by a model, which accounts for three contributions to the profile: a coherent and an incoherent contribution of the first excited state as well as an incoherent contribution of the vibrational ground state (see Fig.~\ref{fig:1D_coherence}c,f). For the incoherent contributions, we assume the corresponding Wannier functions of the ground and first excited states, respectively. We use a minimum coherence length of 6 lattice sites as a criterion for separating the coherent and the incoherent fractions of the excited band population. 
 
Fig.~\ref{fig:1D_coherence}g shows the time evolution of the three contributions extracted from a fit for $V_{hold}=20\,E_r$.
Within the first millisecond, we observe damped oscillations in the coherent and incoherent fractions in the excited band with a frequency of $\omega_\mathrm{osc}=2\pi\times2.6(1)\,$kHz. The origin of these oscillations, which only emerge for a small range of the potential depths along the excitation axis, is not fully understood so far. Possible explanations could be inter- and intra-band interference effects, or the onset of an oscillating superfluid order parameter as predicted for a rapid projection of a Mott-insulating system into the superfluid regime \cite{Altman02,Polkovnikov02}.
For hold times longer than 1\,ms, the oscillations of the coherent and incoherent contributions are damped and reach a steady state from which both populations subsequently decay as discussed in the  previous section. 


\begin{figure}
	\begin{center}
	\includegraphics[width=
\columnwidth]{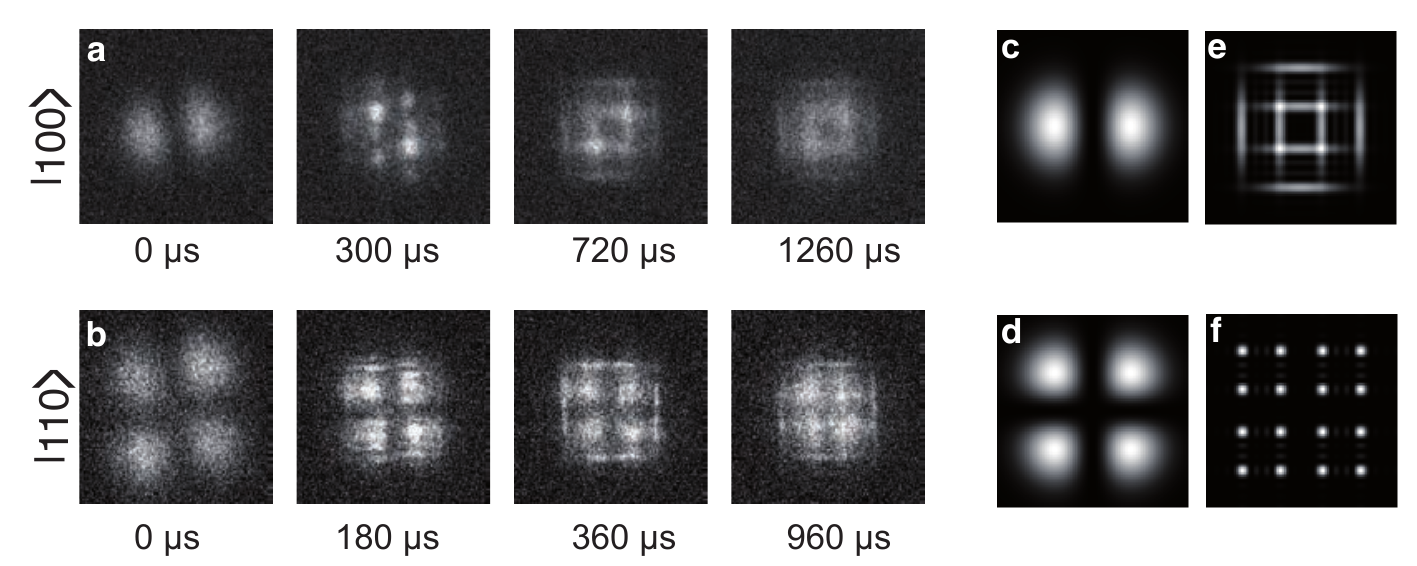}
\end{center}
\caption[]{Time evolution of the measured interference pattern emerging out of the states {\bf (a)} $\left|100\right\rangle$ and {\bf (b)} $\left|110\right\rangle$  after the $x-$ and $y$-axes were ramped down to 17\,$E_{R}$ in 50\,$\rm{\mu}$s after the state preparation while the $z$-axes was kept at 55\,$E_{R}$. In analogy to the 1D case, for zero hold time {\bf (a,b)} the corresponding Wannier product wave functions of $\left|100\right\rangle$ and $\left|110\right\rangle$ {\bf (c+d)} are visible, indicating an initial homogeneous occupation of the first excited Bloch band along {\bf (a)} one and {\bf (b)} two axes. A totally relaxed system in the two cases should yield grid- {\bf (e)} and point-like {\bf (f)} interference patterns.}
\label{fig:2D_Coherence}
\end{figure}

Vibrational excitations in energetically degenerate lattice axes are coupled, if the axes are not aligned perfectly perpendicular. This is the case in our setup. Consequently, for the same sequence as used to observe the emergence of coherence in 1D, an excitation along $\left|100\right\rangle$ can couple to $\left|010\right\rangle$ if the potential along the $y-$axis is ramped after the Raman $\pi$-pulse such that $V_y = V_x$. The ensuing dynamics in 2D can be observed (see Fig.~\ref{fig:2D_Coherence}a) and shows an intriguing time evolution. By a slightly more intricate experimental sequence, the system can also be initially excited to $\left|110\right\rangle$. Here, we also observe a pronounced dynamics (see Fig.~\ref{fig:2D_Coherence}b), where the momentum distribution in both cases finally relaxes to a grid-like structure after approximately 400\,$\rm{\mu}$s hold time. This alludes to a situation where coherence is built up along the two spatial directions independently. The particles thus seem to tunnel preferentially in one direction, rather than along two axes, thus yielding no emerging cross-coherence between the two directions.


In conclusion, we have demonstrated the coherent manipulation of vibrational bands in an optical lattice, thereby realizing a $p$-wave orbital system within a 3D optical lattice. We have measured the stability of atoms in the first excited Bloch band and could observe lifetimes one or two orders of magnitude larger than the characteristic time scales for tunneling. Finally, we have observed the dynamical emergence of coherence in 1D as well as in 2D, where the system relaxes to a non-zero momentum state including an oscillatory dynamical evolution. Due to the long lifetimes, we believe that this system is a promising starting point for the realization of multi-orbital quantum phases with ultracold atoms.


We acknowledge financial support from the DFG and the EU under a Marie Curie excellence grant (QUASICOMBS).


\begin{thebibliography}{20}
\expandafter\ifx\csname natexlab\endcsname\relax\def\natexlab#1{#1}\fi
\expandafter\ifx\csname bibnamefont\endcsname\relax
  \def\bibnamefont#1{#1}\fi
\expandafter\ifx\csname bibfnamefont\endcsname\relax
  \def\bibfnamefont#1{#1}\fi
\expandafter\ifx\csname citenamefont\endcsname\relax
  \def\citenamefont#1{#1}\fi
\expandafter\ifx\csname url\endcsname\relax
  \def\url#1{\texttt{#1}}\fi
\expandafter\ifx\csname urlprefix\endcsname\relax\def\urlprefix{URL }\fi
\providecommand{\bibinfo}[2]{#2}
\providecommand{\eprint}[2][]{\url{#2}}

\bibitem[{\citenamefont{Jaksch and Zoller}(2005)}]{jaksch05a}
\bibinfo{author}{\bibfnamefont{D.}~\bibnamefont{Jaksch}} \bibnamefont{and}
  \bibinfo{author}{\bibfnamefont{P.}~\bibnamefont{Zoller}},
  \bibinfo{journal}{Ann.~Phys.~(N.Y.)} \textbf{\bibinfo{volume}{315}},
  \bibinfo{pages}{52} (\bibinfo{year}{2005}).

\bibitem[{\citenamefont{Bloch}(2005)}]{bloch05a}
\bibinfo{author}{\bibfnamefont{I.}~\bibnamefont{Bloch}},
  \bibinfo{journal}{Nature Phys.} \textbf{\bibinfo{volume}{1}},
  \bibinfo{pages}{23} (\bibinfo{year}{2005}).

\bibitem[{\citenamefont{Isacsson and Girvin}(2005)}]{Girvin05}
\bibinfo{author}{\bibfnamefont{A.}~\bibnamefont{Isacsson}} \bibnamefont{and}
  \bibinfo{author}{\bibfnamefont{S.} \bibnamefont{Girvin}},
  \bibinfo{journal}{Phys.~Rev.~A} \textbf{\bibinfo{volume}{72}},
  \bibinfo{pages}{053604} (\bibinfo{year}{2005}).

\bibitem[{\citenamefont{Kuklov}(2006)}]{Kuklov06}
\bibinfo{author}{\bibfnamefont{A.}~\bibnamefont{Kuklov}},
  \bibinfo{journal}{Phys.~Rev.~Lett.} \textbf{\bibinfo{volume}{97}},
  \bibinfo{pages}{110405} (\bibinfo{year}{2006}).

\bibitem[{\citenamefont{Xu and Fisher}(2007)}]{Xu07}
\bibinfo{author}{\bibfnamefont{C.}~\bibnamefont{Xu}} \bibnamefont{and}
  \bibinfo{author}{\bibfnamefont{M.}~\bibnamefont{Fisher}},
  \bibinfo{journal}{cond--mat/0611620}.

\bibitem[{\citenamefont{Scarola and Sarma}(2005)}]{Scarola05}
\bibinfo{author}{\bibfnamefont{V.}~\bibnamefont{Scarola}} \bibnamefont{and}
  \bibinfo{author}{\bibfnamefont{S.} \bibnamefont{Das Sarma}},
  \bibinfo{journal}{Phys.~Rev.~Lett.} \textbf{\bibinfo{volume}{95}},
  \bibinfo{eid}{033003} (\bibinfo{year}{2005}).

\bibitem[{\citenamefont{Scarola et~al.}(2006)\citenamefont{Scarola, Demler, and
  Sarma}}]{Scarola06}
\bibinfo{author}{\bibfnamefont{V.} \bibnamefont{Scarola}},
  \bibinfo{author}{\bibfnamefont{E.}~\bibnamefont{Demler}}, \bibnamefont{and}
  \bibinfo{author}{\bibfnamefont{S.} \bibnamefont{Das Sarma}},
  \bibinfo{journal}{Phys.~Rev.~A} \textbf{\bibinfo{volume}{73}},
  \bibinfo{eid}{051601} (\bibinfo{year}{2006}).

\bibitem[{\citenamefont{Wu et~al.}(2006)\citenamefont{Wu, Liu, Moore, and
  Sarma}}]{Wu06}
\bibinfo{author}{\bibfnamefont{C.}~\bibnamefont{Wu}},
  \bibinfo{author}{\bibfnamefont{W.}~\bibnamefont{Liu}},
  \bibinfo{author}{\bibfnamefont{J.}~\bibnamefont{Moore}}, \bibnamefont{and}
  \bibinfo{author}{\bibfnamefont{S.} \bibnamefont{Das Sarma}},
  \bibinfo{journal}{Phys.~Rev.~Lett.} \textbf{\bibinfo{volume}{97}},
  \bibinfo{pages}{190406} (\bibinfo{year}{2006}).

\bibitem[{\citenamefont{Wu et~al.}(2007)\citenamefont{Wu, Bergman, Balents, and
  Sarma}}]{Wu07}
\bibinfo{author}{\bibfnamefont{C.}~\bibnamefont{Wu}},
  \bibinfo{author}{\bibfnamefont{D.}~\bibnamefont{Bergman}},
  \bibinfo{author}{\bibfnamefont{L.}~\bibnamefont{Balents}}, \bibnamefont{and}
  \bibinfo{author}{\bibfnamefont{S.} \bibnamefont{Das Sarma}},
  \bibinfo{journal}{cond--mat/0701788}.

\bibitem[{\citenamefont{Leibfried et~al.}(2003)\citenamefont{Leibfried, Blatt,
  Monroe, and Wineland}}]{Wineland03}
\bibinfo{author}{\bibfnamefont{D.}~\bibnamefont{Leibfried}},
  \bibinfo{author}{\bibfnamefont{R.}~\bibnamefont{Blatt}},
  \bibinfo{author}{\bibfnamefont{C.}~\bibnamefont{Monroe}}, \bibnamefont{and}
  \bibinfo{author}{\bibfnamefont{D.}~\bibnamefont{Wineland}},
  \bibinfo{journal}{Rev.~Mod.~Phys,} \textbf{\bibinfo{volume}{75}},
  \bibinfo{pages}{281} (\bibinfo{year}{2003}).

\bibitem[{\citenamefont{Spielman et~al.}(2006)\citenamefont{Spielman, Johnson,
  Huckans, Fertig, Rolston, Phillips, and Porto}}]{Spielmann06}
\bibinfo{author}{\bibfnamefont{I.}~\bibnamefont{Spielman}},
  \bibinfo{author}{\bibfnamefont{P.}~\bibnamefont{Johnson}},
  \bibinfo{author}{\bibfnamefont{J.}~\bibnamefont{Huckans}},
  \bibinfo{author}{\bibfnamefont{C.}~\bibnamefont{Fertig}},
  \bibinfo{author}{\bibfnamefont{S.} \bibnamefont{Rolston}},
  \bibinfo{author}{\bibfnamefont{W.}~\bibnamefont{Phillips}}, \bibnamefont{and}
  \bibinfo{author}{\bibfnamefont{J.V.}~\bibnamefont{Porto}},
  \bibinfo{journal}{Phys.~Rev.~A} \textbf{\bibinfo{volume}{73}},
  \bibinfo{pages}{020702} (\bibinfo{year}{2006}).

\bibitem[{\citenamefont{Liu and Wu}(2006)}]{Liu06}
\bibinfo{author}{\bibfnamefont{W.}~\bibnamefont{Liu}} \bibnamefont{and}
  \bibinfo{author}{\bibfnamefont{C.}~\bibnamefont{Wu}},
  \bibinfo{journal}{Phys.~Rev.~A} \textbf{\bibinfo{volume}{74}},
  \bibinfo{pages}{013607} (\bibinfo{year}{2006}).

\bibitem[{\citenamefont{Chin et~al.}(2006)\citenamefont{Chin, Miller, Liu,
  Stan, Setiawan, Sanner, Xu, and Ketterle}}]{Chin06}
\bibinfo{author}{\bibfnamefont{J.}~\bibnamefont{Chin}},
  \bibinfo{author}{\bibfnamefont{D.}~\bibnamefont{Miller}},
  \bibinfo{author}{\bibfnamefont{Y.}~\bibnamefont{Liu}},
  \bibinfo{author}{\bibfnamefont{C.}~\bibnamefont{Stan}},
  \bibinfo{author}{\bibfnamefont{W.}~\bibnamefont{Setiawan}},
  \bibinfo{author}{\bibfnamefont{C.}~\bibnamefont{Sanner}},
  \bibinfo{author}{\bibfnamefont{K.}~\bibnamefont{Xu}}, \bibnamefont{and}
  \bibinfo{author}{\bibfnamefont{W.}~\bibnamefont{Ketterle}},
  \bibinfo{journal}{Nature} \textbf{\bibinfo{volume}{443}},
  \bibinfo{pages}{961} (\bibinfo{year}{2006}).

\bibitem[{\citenamefont{Köhl et~al.}(2005)\citenamefont{Köhl, Moritz, Stöferle,
  Günter, and Esslinger}}]{Koehl2005a}
\bibinfo{author}{\bibfnamefont{M.}~\bibnamefont{Köhl}},
  \bibinfo{author}{\bibfnamefont{H.}~\bibnamefont{Moritz}},
  \bibinfo{author}{\bibfnamefont{T.}~\bibnamefont{Stöferle}},
  \bibinfo{author}{\bibfnamefont{K.}~\bibnamefont{Günter}}, \bibnamefont{and}
  \bibinfo{author}{\bibfnamefont{T.}~\bibnamefont{Esslinger}},
  \bibinfo{journal}{Phys.~Rev.~Lett.} \textbf{\bibinfo{volume}{94}},
  \bibinfo{pages}{080403} (\bibinfo{year}{2005}).

\bibitem[{\citenamefont{Greiner et~al.}(2001)\citenamefont{Greiner, Bloch,
  Mandel, Hänsch, and Esslinger}}]{Greiner01f}
\bibinfo{author}{\bibfnamefont{M.}~\bibnamefont{Greiner}},
  \bibinfo{author}{\bibfnamefont{I.}~\bibnamefont{Bloch}},
  \bibinfo{author}{\bibfnamefont{O.}~\bibnamefont{Mandel}},
  \bibinfo{author}{\bibfnamefont{T.~W.} \bibnamefont{Hänsch}},
  \bibnamefont{and}
  \bibinfo{author}{\bibfnamefont{T.}~\bibnamefont{Esslinger}},
  \bibinfo{journal}{Phys.~Rev.~Lett.} \textbf{\bibinfo{volume}{87}},
  \bibinfo{pages}{160405} (\bibinfo{year}{2001}).

\bibitem[{\citenamefont{Kastberg et~al.}(1995)\citenamefont{Kastberg, Phillips,
  Rolston, and Spreeuw}}]{Kastberg95}
\bibinfo{author}{\bibfnamefont{A.}~\bibnamefont{Kastberg}},
  \bibinfo{author}{\bibfnamefont{W.}~\bibnamefont{Phillips}},
  \bibinfo{author}{\bibfnamefont{S.} \bibnamefont{Rolston}},
  \bibnamefont{and} \bibinfo{author}{\bibfnamefont{R.}~\bibnamefont{Spreeuw}},
  \bibinfo{journal}{Phys.~Rev.~Lett.} \textbf{\bibinfo{volume}{74}},
  \bibinfo{pages}{1542} (\bibinfo{year}{1995}).

\bibitem[{\citenamefont{Fölling et~al.}(2006)\citenamefont{Fölling, Widera,
  Müller, Gerbier, and Bloch}}]{Foelling06}
\bibinfo{author}{\bibfnamefont{S.}~\bibnamefont{Fölling}},
  \bibinfo{author}{\bibfnamefont{A.}~\bibnamefont{Widera}},
  \bibinfo{author}{\bibfnamefont{T.}~\bibnamefont{Müller}},
  \bibinfo{author}{\bibfnamefont{F.}~\bibnamefont{Gerbier}}, \bibnamefont{and}
  \bibinfo{author}{\bibfnamefont{I.}~\bibnamefont{Bloch}},
  \bibinfo{journal}{Phys.~Rev.~Lett.} \textbf{\bibinfo{volume}{97}},
  \bibinfo{pages}{060403} (\bibinfo{year}{2006}).

\bibitem[{\citenamefont{Altman and Auerbach}(2002)}]{Altman02}
\bibinfo{author}{\bibfnamefont{E.}~\bibnamefont{Altman}} \bibnamefont{and}
  \bibinfo{author}{\bibfnamefont{A.}~\bibnamefont{Auerbach}},
  \bibinfo{journal}{Phys.~Rev.~Lett.} \textbf{\bibinfo{volume}{89}},
  \bibinfo{pages}{250404} (\bibinfo{year}{2002}).

\bibitem[{\citenamefont{Polkovnikov et~al.}(2002)\citenamefont{Polkovnikov,
  Sachdev, and Girvin}}]{Polkovnikov02}
\bibinfo{author}{\bibfnamefont{A.}~\bibnamefont{Polkovnikov}},
  \bibinfo{author}{\bibfnamefont{S.}~\bibnamefont{Sachdev}}, \bibnamefont{and}
  \bibinfo{author}{\bibfnamefont{S.}~\bibnamefont{Girvin}},
  \bibinfo{journal}{Phys.~Rev.~A} \textbf{\bibinfo{volume}{66}},
  \bibinfo{pages}{053607} (\bibinfo{year}{2002}).

\end{thebibliography}

\end{document}